\documentclass[aps,pra,showpacs,twocolumn,floatfix]{revtex4}
\usepackage{graphicx}
\usepackage{dcolumn}

\begin{document}

\title{Relativistic many-body calculations of transition rates
from core-excited states in sodiumlike ions}

\author{U.I. Safronova}
\email{usafrono@nd.edu}
\author{W. R. Johnson}
\email{johnson@nd.edu} \homepage{www.nd.edu/~johnson}
\author{M.S. Safronova}
 \altaffiliation[Current address: ]{Electron and Optical Physics
Division, National Institute of Standards and Technology,
Gaithersburg, MD, 20899-8410}
\affiliation{%
Department of Physics, 225 Nieuwland Science Hall\\
University of Notre Dame, Notre Dame, IN 46566}%

\author{J. R. Albritton}
\affiliation{Lawrence Livermore National Laboratory, PO Box 808,
Livermore, CA 94551}

\date{\today}

\begin{abstract}

Rates and line strengths are calculated
for the $%
2s^22p^53l3l'$ - $2s^22p^63l''$ and $2s2p^63l3l'$ - $2s^22p^63l''$
electric-dipole (E1) transitions
in Na-like ions with
nuclear charges ranging from $Z$ = 14 to 100. Relativistic
many-body perturbation theory (RMBPT), including the
Breit interaction, is used to evaluate retarded E1 matrix elements
in length and velocity forms. The calculations start from a
$1s^22s^22p^6$ Dirac-Fock potential. First-order RMBPT is used to
obtain intermediate coupling coefficients and second-order RMBPT
is used to calculate transition matrix elements. A detailed
discussion of the various contributions to  dipole matrix
elements is given for sodiumlike copper
($Z$ = 29). Transition energies used in the calculation of
 transition rates are from second-order
RMBPT. Trends of transition rates as functions of $Z$ are
shown graphically for selected transitions.
\end{abstract}



\maketitle

\section{Introduction}

Transitions from $2s^22p^53l3l^{\prime}$ and $2s2p^63l3l^{\prime}$
states to the ground ($2s^22p^63s$) or singly-excited ($2s^22p^63p$ and $%
2s^22p^63d$) states form satellite lines to the bright electric-dipole (E1)
lines created by transitions from $2s^22p^53l$ and $2s2p^63l$ states to
the ground state ($2s^22p^6$) in Ne-like ions. These core-excited $%
2s^22p^53l3l^{\prime}$ and $2s2p^63l3l^{\prime}$ states (often
called doubly-excited states) in sodiumlike ions have been
studied extensively both experimentally and theoretically over the
past 20-30 years.

Transition rates and oscillator strengths for Na-like ions have been
calculated using multi-configuration Dirac-Fock (MCDF) \cite{m2,m3} and
multi-configuration Hartree-Fock (MCHF) \cite{m1,bruch,safr}  methods. Recently,
R-matrix calculations of electron-impact collision strengths for excitations
from the inner $L$-shell into doubly excited states of Fe$^{15+}$ was
presented by \citet{baut}. Energies of
$2s^22p^63l$ and $2s^22p^53l3l^{\prime}$ states were calculated in that paper
using the SUPERSTRUCTURE code \cite{eiss74}. It was shown in \cite{baut}
that disagreement between calculated data and data recommended by 
\citet{sugar} and \citet{shirai} ranges from 0.5\%  to 5\%.

Experimentally, the core-excited $2s^22p^53l3l^{\prime}$ and $%
2s2p^63l3l^{\prime}$ states were studied by the beam-foil technique
\cite{a2a,a2}, photo-emission
\cite{a1,m93,m94,m5,m5a,x93,b86,b95,brown,phil}, and Auger
spectroscopy \cite{a3,a3a,m4,alla,a5}. Strong line radiation involving $n$ = 3
to $n$ = 2 transitions in Ne-like ions together with satellite lines of
Na-like ions were observed from laser-produced plasmas
\cite{a1,m93,m94,m5,m5a}, X-pinch, \cite{x93}, tokamaks \cite{b86,b95}, EBIT 
\cite{brown}, and solar flares \cite{phil}. The identification
of measured spectral lines was based on theoretical calculations
carried out primarily using the MCDF method with Cowan's code \cite{cowan}.

In this paper, we present a comprehensive set of calculations 
for $2s^22p^53l3l'$ - $2s^22p^63l''$ and $2s2p^63l3l'$ - $2s^22p^63l''$
transitions to compare with previous calculations and experiments.
Our aim is to provide benchmark values
for the entire Na isoelectronic sequence. The large number of
possible transitions have made experimental identification
difficult. Experimental verifications should become simpler and
more reliable using this  more accurate set of calculations.

Relativistic many-body perturbation theory (RMBPT) is
used here to determine matrix elements and transition
rates for allowed and forbidden electric-dipole transitions between the
odd-parity core-excited states ($2s^22p^53s^2$ + $2s^22p^53p^2$ + $%
2s^22p^53d^2$ + $2s^22p^53s3d$ + $2s2p^63s3p$ + $2s2p^63p3d$) and the ground
state ($2s^22p^63s$) together with the two singly-excited states ($%
2s^22p^63d $) and the even-parity core-excited states ($2s^22p^53s3p$ + $%
2s^22p^53p3d$ + $2s2p^63s^2$ + $2s2p^63p^2$ + $2s2p^63d^2$ +
$2s2p^63s3d$) and the two singly-excited states ($2s^22p^63p$) in
Na-like ions with nuclear charges ranging from $Z$ = 14 to 100.
Retarded E1 matrix elements are evaluated in both length and
velocity forms. These calculations start from a
Ne-like core Dirac-Fock (DF) potential. First-order
perturbation theory is used to obtain intermediate coupling
coefficients and second-order RMBPT is used to determine
transition matrix elements. The energies used in the
calculation of  transition rates are
obtained from second-order RMBPT.

\section{Method}

In this section, we discuss relativistic RMBPT
for first- and second-order transition matrix elements for atomic systems
with two valence electrons and one hole. Details of the RMBPT method for
calculation of radiative transition rates for systems with one valence
electron and one hole were presented for Ne-like and Ni-like ions in \cite{ni,ne}.
Here, we  follow the pattern of the corresponding calculation in
Refs.~\cite{ni,ne} but limit our discussion to the model space
and the first- and second-order 
particle-particle-hole  diagram contributions in Na-like ions.

\subsection{Model space}

For Na-like ions with two electrons above the Ne-like ($1s^{2}2s^{2}2p_{1/2}^{2}2p_{3/2}^{4}$)
core and one hole in the core, the model space is formed from
particle-particle-hole states of the type
$a_{v}^{\dagger}a_{w}^{\dagger}a_{a}| 0 \rangle$, where $|0 \rangle$ is
the core state function.
Indices $v$ and $w$ designate valence electrons and
$a$ designates a core electron.
For our study of low-lying
$3l3l^{\prime}2l^{-1}$ states of Na-like ions, the index $a$ ranges over
$2s$, $2p_{1/2}$, and $2p_{3/2}$, while
$v$ and $w$ range over $3s$, $3p_{1/2}$, $3p_{3/2}$, $3d_{3/2}$, and $%
3d_{5/2}$. To obtain orthonormal model states, we consider the
coupled states $(vwa)$ defined by
\begin{equation}\label{eq1}
\Psi (QJM)=N(Q)\sum \langle vw|K_{12}\rangle
\langle K_{12}a|K\rangle a_{v}^{\dagger }a_{w}^{\dagger
}a_{a}|0\rangle \,,
\end{equation}
where  $Q$ describes a particle-particle-hole state with quantum numbers $%
n_v\kappa_vn_w\kappa_w[J_{12}] n_a\kappa_a$ and intermediate
momentum $J_{12}$. We use the notation $K_i=\{J_i, M_i\}$ and $v =\{j_v,
m_v\}$. The sum in Eq.(\ref{eq1}) is over magnetic quantum numbers $m_v$, $%
m_w$, $m_a$, and $M_{12}$. The quantity $\langle K_1 K_2|K_3 \rangle$ is a
Clebsch-Gordan coefficient:
\begin{equation}  \label{eq2}
\langle K_1 K_2| K_3 \rangle=(-1)^{J_1-J_2+M_3} \sqrt{[J_3]} \left(
\begin{array}{ccc}
J_1 & J_2 & J_3 \\
M_1 & M_2 & -M_3
\end{array}
\right) \,,
\end{equation}
where $[J]=2J+1$.
Combining two $n = 3$ particles  with possible intermediate momenta
and $n = 2$ hole orbitals in sodium, we obtain 121 odd-parity states
with  $J = 1/2\, \cdots\, 11/2$ and 
116 even-parity states with $J = 1/2\, \cdots\, 11/2$.
The distribution of the 237 states in the model space is
found in Table I of the accompanying EPAPS document \cite{EPAPS}.
Instead of using the $3l^{\prime}3l^{\prime%
\prime}[J_{1}]2l^{-1}(J)$ designations, we use simpler designations $%
3l^{\prime}3l^{\prime\prime}[J_{1}]2l(J)$  in
the tables and text below.

\subsection{Dipole matrix element}

The first- and second-order reduced E1 matrix elements $Z^{(1)}$,  and $Z^{(2)}$,
and the second-order Breit correction to the reduced E1 matrix element $B^{(2)}$ 
for a
transition between the uncoupled particle-particle-hole
state $\Psi (QJM)$ of Eq.~(\ref{eq1}) and the one-particle state $%
a_{x}^{\dagger}|0\rangle $ are given in Appendix.

\begin{figure*}
\centerline{\includegraphics[scale=0.35]{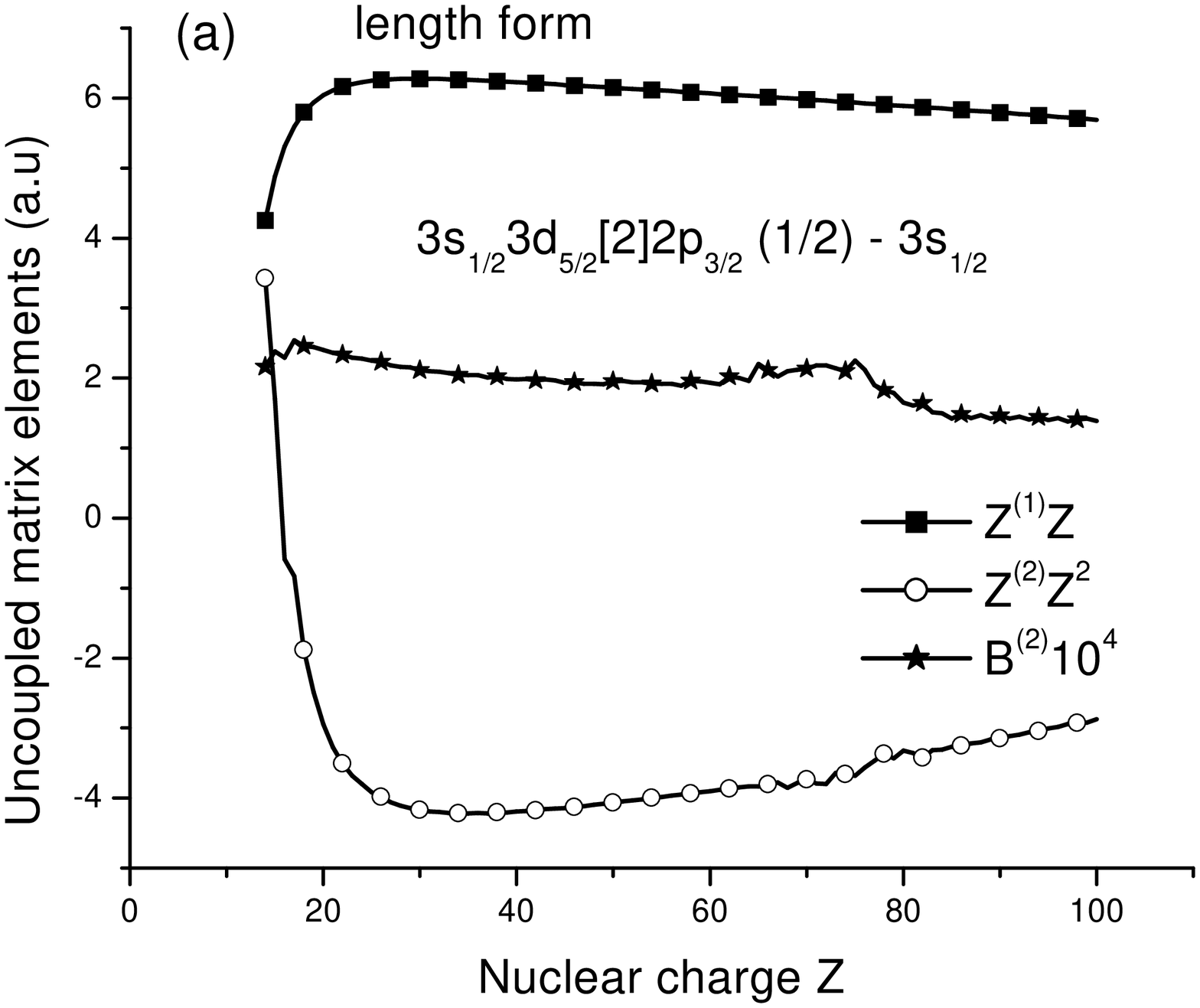}
            \includegraphics[scale=0.35]{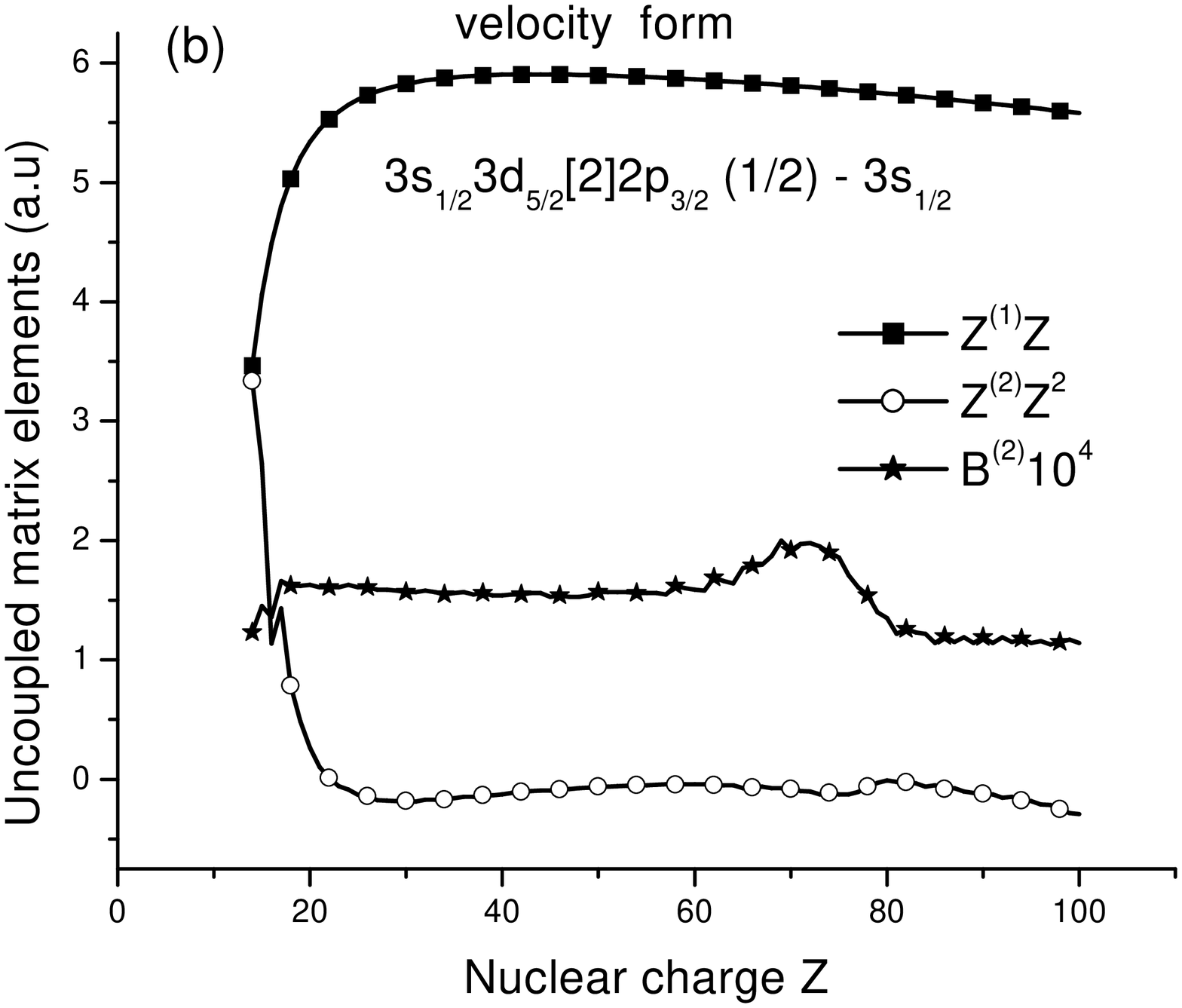}}
           \caption{Uncoupled matrix element for
$3s3d_{5/2}[2]2p_{3/2}\ (1/2)$--$3s$ transition
calculated in length and velocity forms in Na-like ions.}
\label{fig1}
\end{figure*}

The uncoupled reduced matrix elements are calculated in both
length and velocity gauges. Differences between length and
velocity forms are illustrated for the uncoupled
[$3s3d_{5/2}[2]2p_{3/2}\ (1/2)$ - $3s$]  matrix element
in panels (a) and (b) of Fig.~\ref{fig1}. In the high-$Z$ limit, 
$Z^{(1)}$ is proportional to $1/Z$, 
 $Z^{(2)} $  is proportional to $1/Z^2$, and
 $B^{(2)} $ is independent of $Z$ (see \cite {be-tr}). Taking into account this
$Z$-dependence, we plot $Z^{(1)}\times Z$, $Z^{(2)}\times Z^2$, and
$B^{(2)}\times 10^4$ in the figure.  The contribution of the second-order
matrix elements $Z^{(2)}$ is seen to be much larger in length form. 
Differences
between results in length and velocity  forms shown in
Fig.~\ref{fig1} are precisely compensated by ``derivative terms'' $P^{({\rm
derv})}$, as shown later.

\begin{table*}
\caption{Uncoupled reduced matrix elements in length $L$ and velocity $V$ forms for
transitions between the selected odd-parity core-excited states
with $J$ = 1/2 and the ground $3s$ and singly-excited $3d_{3/2}$ states in Cu$^{18+}$ ion.}
\begin{ruledtabular}\begin{tabular}{lrrrrrrrr}
\multicolumn{1}{c}{$3lj3l'j'(J_1)2l''j''$}&
\multicolumn{1}{c}{$Z^{(1)}_L$} &
\multicolumn{1}{c}{$Z^{(1)}_V$} &
\multicolumn{1}{c}{$Z^{(2)}_L$} &
\multicolumn{1}{c}{$Z^{(2)}_V$} &
\multicolumn{1}{c}{$B^{(2)}_L$} &
\multicolumn{1}{c}{$B^{(2)}_V$} &
\multicolumn{1}{c}{$P^{(\rm derv)}_L$} &
\multicolumn{1}{c}{$P^{(\rm derv)}_V$}\\[0.25pc]
\hline
\multicolumn{9}{c} {[$3lj3l'j'[J_1]2l''j''\ (1/2)$-- $3s_{1/2}$] transitions} \\
\hline
$3s_{1/2} 3s_{1/2}[0]2p_{1/2}$& -0.040775&-0.037586&-0.003606&-0.002838&-0.000116&-0.000104&-0.040564& 0.000139\\
$3s_{1/2} 3d_{3/2}[1]2p_{3/2}$& -0.057161&-0.052902& 0.001364& 0.000191&-0.000070&-0.000108&-0.057130&-0.000115\\
$3p_{1/2} 3p_{3/2}[1]2p_{3/2}$&  0.000000& 0.000000&-0.000160&-0.000141&-0.000002&-0.000003& 0.000000& 0.000000\\
$3p_{1/2} 3p_{3/2}[2]2p_{3/2}$&  0.000000& 0.000000&-0.001048&-0.001313& 0.000002& 0.000003& 0.000000& 0.000000\\
$3p_{3/2} 3p_{3/2}[2]2p_{3/2}$&  0.000000& 0.000000&-0.000889&-0.001067&-0.000002&-0.000001& 0.000000& 0.000000\\
\hline
\multicolumn{9}{c} {[$3lj3l'j'[J_1]2l''j''\ (1/2)$-- $3d_{3/2}$] transitions} \\
\hline
$3s_{1/2} 3s_{1/2}[0]2p_{1/2}$&  0.000000& 0.000000& 0.000388&-0.000293& 0.000006& 0.000003& 0.000000& 0.000000\\
$3s_{1/2} 3d_{3/2}[1]2p_{3/2}$&  0.012526& 0.011536& 0.001077& 0.000835& 0.000027& 0.000013& 0.012483&-0.000003\\
$3p_{1/2} 3p_{3/2}[1]2p_{3/2}$&  0.000000& 0.000000& 0.000205& 0.000602&-0.000001&-0.000001& 0.000000& 0.000000\\
$3p_{1/2} 3p_{3/2}[2]2p_{3/2}$&  0.000000& 0.000000& 0.000720&-0.000132&-0.000002& 0.000001& 0.000000& 0.000000\\
$3p_{3/2} 3p_{3/2}[2]2p_{3/2}$&  0.000000& 0.000000& 0.000671& 0.000436& 0.000004& 0.000002& 0.000000& 0.000000\\
\end{tabular}\end{ruledtabular}
\label{tab-uncl}
\end{table*}

\begin{table*}
\caption{Wavelengths ($\lambda$ in \AA) and transition rates ($A_r$ in $s^{-1}$)  for transitions from
core-excited states $QJ$ ($Q = 3lj3l'j'[J_1]2l''j''$, $J=1/2$ ) to the
ground  state in Na-like ions.
 Comparison with theoretical data obtained by using YODA code from
Ref.~\protect\cite{m3}.
 Numbers in brackets represent powers of 10.}
\begin{ruledtabular}\begin{tabular}{lrrrrrrrr}
\multicolumn{1}{c} {Q} &
\multicolumn{4}{c} {$Z$ = 26} &
\multicolumn{4}{c} {$Z$ = 54} \\
\multicolumn{1}{c} {} & \multicolumn{1}{c} {$\lambda_{\rm
RMBPT}$}& \multicolumn{1}{c} {$\lambda_{\rm YODA}$}&
\multicolumn{1}{c} {$A_{r}^{\rm RMBPT}$}& \multicolumn{1}{c}
{$A_{r}^{\rm YODA}$} & \multicolumn{1}{c} {$\lambda_{\rm RMBPT}$}&
\multicolumn{1}{c} {$\lambda_{\rm YODA}$}& \multicolumn{1}{c}
{$A_{r}^{\rm RMBPT}$}&
\multicolumn{1}{c} {$A_{r}^{\rm YODA}$} \\
\hline
$3s_{1/2}3s_{1/2}[0] 2p_{1/2}$& 17.0747& 17.1207& 8.096[11]& 8.27[11]& 2.7865&  2.7884&  2.507[12]& 2.62[12]\\
$3s_{1/2}3d_{3/2}[1] 2p_{3/2}$& 15.9418& 15.9883& 9.449[10]& 1.10[11]& 2.7731&  2.7747&  9.345[10]& 6.70[10]\\
$3p_{1/2}3p_{1/2}[0] 2p_{1/2}$& 15.6327& 15.6671& 2.578[09]& 2.67[09]& 2.7355&  2.7366&  2.625[14]& 2.50[14]\\
$3s_{1/2}3d_{3/2}[2] 2p_{3/2}$& 15.5711& 15.5958& 4.600[11]& 5.21[11]& 2.7241&  2.7251&  1.724[14]& 2.07[14]\\
$3s_{1/2}3d_{5/2}[2] 2p_{3/2}$& 15.5029& 15.5193& 3.776[12]& 4.02[12]& 2.6271&  2.6276&  4.846[11]& 6.35[11]\\
$3p_{1/2}3p_{3/2}[1] 2p_{1/2}$& 15.3681& 15.3884& 1.465[11]& 1.56[11]& 2.5927&  2.5939&  8.932[10]& 1.06[12]\\
$3p_{3/2}3p_{3/2}[0] 2p_{1/2}$& 15.3672& 15.3558& 9.425[08]& 7.81[09]& 2.5925&  2.5938&  9.882[11]& 6.29[10]\\
$3s_{1/2}3d_{3/2}[1] 2p_{1/2}$& 15.2148& 15.2174& 2.240[13]& 2.65[13]& 2.5734&  2.5748&  9.387[11]& 6.72[11]\\
$3s_{1/2}3p_{1/2}[0] 2s_{1/2}$& 14.0969& 14.1212& 2.826[10]& 2.80[10]& 2.5572&  2.5580&  5.078[13]& 6.02[13]\\
$3d_{5/2}3d_{5/2}[0] 2p_{1/2}$& 13.6959& 13.6884& 9.156[07]& 1.67[09]& 2.3876&  2.3881&  2.952[08]& 1.43[10]\\
\end{tabular}
\end{ruledtabular}
\label{tab-nils}
\end{table*}

\subsection{Dipole matrix elements in Cu$^{18+}$ }
In Table~\ref{tab-uncl}, we list values of {\em uncoupled} first-
and second-order dipole matrix elements $Z^{(1)}$, $Z^{(2)}$,
$B^{(2)}$, together with derivative terms $P^{({\rm derv})}$ for
Na-like copper, $Z$ = 29. For simplicity, we only list values for
selected dipole transitions between  odd-parity  states with $J$ = 1/2
and the ground $3s$ and excited $3d_{3/2}$ states. Uncoupled matrix
elements for other transitions in Na-like copper
are given in Table II of the accompanying EPAPS document \cite{EPAPS}. The
derivative terms shown in Table~\ref{tab-uncl} arise because
transition amplitudes depend on energy, and the transition energy
changes order-by-order in RMBPT calculations. Both length ($L$)
and velocity ($V$) forms are given for the matrix elements. We find that
the first-order matrix elements $Z^{(1)}_L$ and
$Z^{(1)}_V$ differ by 10\%; the $L$ - $V$ differences between
second-order matrix elements are much larger for some transitions.
The term  $P^{({\rm derv)}}$ in length form
almost equals $Z^{(1)}$ in length form but  $P^{({\rm
derv)}}$ in velocity form is smaller than $Z^{(1)}$
in velocity form by three to four orders of magnitude.

 Although we use an
intermediate-coupling scheme, it is nevertheless convenient to
label the physical states using the $LS$ scheme. Length and velocity
 forms of  coupled matrix elements
  differ  only in the fourth or fifth digits.
These $L$ - $V$ differences arise because we start our RMBPT
calculations using a non-local Dirac-Fock  potential.  If we
were to replace the DF potential by a local potential, the
differences would disappear completely.  Removing the
second-order contribution increases $L$ - $V$ differences by a
factor of 10.
Values of {\em coupled} reduced matrix elements in length and
velocity forms are given in Table III of the accompanying EPAPS document \cite{EPAPS}.
Theoretical wavelengths ${\lambda}$ and transition
probabilities $A_r$ for selected transitions in Na-like from
$Z$ = 26 up to $Z$ = 30 are given in Table IV of \cite{EPAPS}.

\begin{table}
\caption{Wavelengths ($\lambda$ in \AA) and transition rates ($A_r$ in $s^{-1}$)for transitions between
core-excited states  $3l3l'(L_1S_1)2p\ LSJ$  and singly-excited
states $3l\ L'S'J'$  in Na-like ions.
 Comparison with experimental data ($\lambda_{\rm expt}$) from
Ref.~\protect\cite{a1}.
 Numbers in brackets represent powers of 10.}
\begin{ruledtabular}\begin{tabular}{llrrr}
\multicolumn{1}{c} {Upper level} &
\multicolumn{1}{c} {Low level}&
\multicolumn{3}{c} {$Z$ = 26} \\
\multicolumn{2}{c} {} &
\multicolumn{1}{c} {$\lambda_{\rm RMBPT}$} &
\multicolumn{1}{c} {$\lambda_{\rm expt}$}&
\multicolumn{1}{c} {$A_r$} \\
\hline
$3s3p(^3P)2p\ ^2S_{1/2}$&$3p\ ^2P_{3/2}$&    16.813&   16.821 &9.189[10]\\
$3s3d(^3D)2p\ ^2P_{3/2}$&$3d\ ^2D_{3/2}$&    16.839&   18.834 &2.942[11]\\
$3s3p(^3P)2p\ ^2D_{3/2}$&$3p\ ^2P_{1/2}$&    16.883&   16.899 &9.942[10]\\
$3s3p(^3P)2p\ ^2P_{1/2}$&$3p\ ^2P_{1/2}$&    16.939&   16.937 &2.782[10]\\
$3s3d(^3D)2p\ ^2P_{1/2}$&$3d\ ^2D_{3/2}$&    16.958&   16.952 &4.962[11]\\
$3s3p(^3P)2p\ ^2P_{1/2}$&$3p\ ^2P_{3/2}$&    16.999&   16.993 &7.880[11]\\
$3s3d(^1D)2p\ ^2P_{3/2}$&$3d\ ^2D_{5/2}$&    17.037&   17.029 &5.370[11]\\
$3p3p(^3P)2p\ ^2P_{3/2}$&$3d\ ^2D_{3/2}$&    17.098&   17.094 &3.901[11]\\
$3s3d(^1D)2p\ ^2D_{5/2}$&$3d\ ^2D_{3/2}$&    17.131&   17.129 &4.887[11]\\
$3s3d(^3D)2p\ ^4F_{3/2}$&$3d\ ^2D_{3/2}$&    17.168&   17.166 &4.273[11]\\
$3s3p(^1P)2p\ ^2P_{3/2}$&$3p\ ^2P_{3/2}$&    17.199&   17.208 &5.081[11]\\
$3s3p(^3P)2p\ ^4P_{3/2}$&$3p\ ^2P_{3/2}$&    17.242&   17.244 &1.972[11]\\
$3s3p(^3P)2p\ ^4D_{1/2}$&$3p\ ^2P_{3/2}$&    17.307&   17.297 &1.118[11]\\
$3s3d(^3D)2p\ ^2P_{3/2}$&$3d\ ^2D_{5/2}$&    17.353&   17.344 &7.288[11]\\
$3s3s(^1S)2p\ ^2P_{3/2}$&$3s\ ^2S_{1/2}$&    17.374&   17.368 &7.925[11]\\
$3p3p(^3P)2p\ ^2D_{5/2}$&$3d\ ^2D_{3/2}$&    17.394&   17.398 &3.118[11]\\
$3s3d(^1D)2p\ ^2P_{1/2}$&$3d\ ^2D_{3/2}$&    17.401&   17.407 &6.364[11]\\
$3s3p(^1P)2p\ ^2S_{1/2}$&$3p\ ^2P_{3/2}$&    17.454&   17.451 &7.665[11]\\
$3s3p(^1P)2p\ ^2P_{1/2}$&$3p\ ^2P_{1/2}$&    17.484&   17.471 &5.848[11]\\
$3s3d(^3D)2p\ ^4F_{7/2}$&$3d\ ^2D_{5/2}$&    17.493&   17.499 &1.394[11]\\
$3s3p(^1P)2p\ ^2P_{1/2}$&$3p\ ^2P_{3/2}$&    17.548&   17.542 &1.810[10]\\
$3p3p(^3P)2p\ ^4D_{1/2}$&$3d\ ^2D_{3/2}$&    17.596&   17.596 &6.959[10]\\
$3s3p(^3P)2p\ ^4S_{3/2}$&$3p\ ^2P_{1/2}$&    17.611&   17.623 &2.480[09]\\
$3s3p(^3P)2p\ ^4S_{3/2}$&$3p\ ^2P_{3/2}$&    17.677&   17.660 &1.185[10]\\
$3p3p(^3P)2p\ ^4D_{5/2}$&$3d\ ^2D_{3/2}$&    17.736&   17.734 &1.366[10]\\
$3p3p(^1D)2p\ ^2D_{5/2}$&$3d\ ^2D_{5/2}$&    17.763&   17.787 &1.443[10]\\
$3p3p(^3P)2p\ ^4P_{3/2}$&$3d\ ^2D_{5/2}$&    17.821&   17.821 &2.537[10]\\
$3p3p(^3P)2p\ ^4P_{3/2}$&$3d\ ^2D_{5/2}$&    17.882&   17.901 &1.892[10]\\
\end{tabular}\end{ruledtabular}
\label{tab-fe}
\end{table}

\begin{table*}
\caption{Wavelengths ($\lambda$ in \AA) and transition rates ($A_r$ in $s^{-1}$) for transitions between
core-excited states  $3l3l'(L_1S_1)2p\ LSJ$  and singly-excited
states $3l\ L'S'J'$  in Na-like ions.
 Comparison with experimental data ($\lambda_{\rm expt}$) from
Ref.~\protect\cite{m94} ($a$) and Ref.~\protect\cite{brown} ($b$).  Numbers in brackets represent powers of 10.}
\begin{ruledtabular}\begin{tabular}{llllrrrr}
\multicolumn{1}{c} {Upper level} & \multicolumn{1}{c} {Low level}
& \multicolumn{3}{c} {$Z$ = 26} & \multicolumn{3}{c} {$Z$ = 27} \\
\multicolumn{2}{c} {} & \multicolumn{1}{c} {$\lambda_{\rm RMBPT}$}&
\multicolumn{1}{c} {$\lambda_{\rm expt}$}& \multicolumn{1}{c}
{$A_r$} & \multicolumn{1}{c} {$\lambda_{\rm RMBPT}$} &
\multicolumn{1}{c} {$\lambda_{\rm expt}$}&
\multicolumn{1}{c} {$A_r$} \\
\hline
$3p3d(^1S)2p\ ^2D_{3/2}$&$  3p\ ^2P_{1/2}$& 15.057&           &5.532[11]&  13.667& 13.667$^a$&7.187[11]\\
$3p3d(^1S)2p\ ^2D_{5/2}$&$  3p\ ^2P_{3/2}$& 15.090&           &1.272[12]&  13.704& 13.707$^a$&2.067[12]\\
$3d3d(^1S)2p\ ^2P_{1/2}$&$  3d\ ^2D_{3/2}$& 15.092& 15.093$^a$&5.375[12]&  13.695&          &6.534[12]\\
$3s3d(^3D)2p\ ^2P_{3/2}$&$  3s\ ^2S_{1/2}$& 15.119& 15.115$^b$&8.210[12]&  13.718&          &1.008[13]\\
$3p3d(^3D)2p\ ^2S_{1/2}$&$  3p\ ^2P_{1/2}$& 15.150& 15.142$^a$&8.704[10]&  13.744&          &1.996[11]\\
$3p3d(^1S)2p\ ^2D_{3/2}$&$  3p\ ^2P_{3/2}$& 15.166&           &3.835[12]&  13.775& 13.773$^a$&4.836[12]\\
$3d3d(^3P)2p\ ^2P_{3/2}$&$  3d\ ^2D_{3/2}$& 15.181& 15.182$^a$&9.080[12]&  13.774&          &1.146[13]\\
$3s3d(^3D)2p\ ^2P_{1/2}$&$  3s\ ^2S_{1/2}$& 15.215& 15.208$^b$&2.240[13]&  13.805&          &2.723[13]\\
$3p3d(^3P)2p\ ^2P_{1/2}$&$  3p\ ^2P_{1/2}$& 15.237&           &1.430[13]&  13.823& 13.823$^a$&1.699[13]\\
$3s3d(^1D)2p\ ^2P_{3/2}$&$  3s\ ^2S_{1/2}$& 15.272& 15.26$^b$ &1.340[13]&  13.855&          &1.614[13]\\
$3s3d(^1D)2p\ ^2P_{3/2}$&$  3s\ ^2S_{1/2}$& 15.272& 15.280$^a$&1.340[13]&  13.855&          &1.614[13]\\
$3p3d(^3P)2p\ ^2P_{1/2}$&$  3p\ ^2P_{3/2}$& 15.286& 15.289$^a$&1.536[11]&  13.872&          &9.115[10]\\
$3d3d(^3F)2p\ ^2F_{5/2}$&$  3d\ ^2D_{5/2}$& 15.361& 15.360$^a$&1.002[12]&  13.930&          &1.691[12]\\
$3d3d(^3F)2p\ ^2D_{5/2}$&$  3d\ ^2D_{3/2}$& 15.369&           &6.618[12]&  13.943& 13.940$^a$&9.585[12]\\
$3p3d(^3D)2p\ ^4F_{3/2}$&$  3p\ ^2P_{3/2}$& 15.429&           &3.019[09]&  14.014& 14.015$^a$&1.919[09]\\
$3s3d(^3D)2p\ ^4D_{3/2}$&$  3s\ ^2S_{1/2}$& 15.435&           &1.742[12]&  14.004& 14.000$^a$&2.880[12]\\
$3p3d(^1D)2p\ ^2D_{3/2}$&$  3p\ ^2P_{1/2}$& 15.510&           &3.134[11]&  14.079& 14.078$^a$&4.868[11]\\
$3s3d(^3D)2p\ ^2P_{3/2}$&$  3s\ ^2S_{1/2}$& 15.525&           &2.997[12]&  14.091& 14.093$^a$&3.817[12]\\
$3d3d(^3F)2p\ ^4F_{3/2}$&$  3d\ ^2D_{5/2}$& 15.561& 15.568$^a$&1.746[11]&  14.129&          &2.141[11]\\
$3d3d(^1D)2p\ ^2F_{7/2}$&$  3d\ ^2D_{5/2}$& 15.573&           &6.467[11]&  14.143& 14.148$^a$&7.639[11]\\
\end{tabular}\end{ruledtabular}
\label{tab-com1}
\end{table*}

\begin{table*}
\caption{Wavelengths ($\lambda$ in \AA) and transition rates ($A_r$ in $s^{-1}$) for transitions between
core-excited states  $3l3l'(L_1S_1)2p\ LSJ$  and singly-excited
states $3l\ L'S'J'$  in Na-like ions.
 Comparison with experimental data ($\lambda_{\rm expt}$) from
Ref.~\protect\cite{m94}.
 Numbers in brackets represent powers of 10.}
\begin{ruledtabular}\begin{tabular}{llrrrrrrrrr}
\multicolumn{1}{c} {Upper level} &
\multicolumn{1}{c} {Low level}&
\multicolumn{3}{c} {$Z$ = 29} &
\multicolumn{3}{c} {$Z$ = 28} &
\multicolumn{3}{c} {$Z$ = 30} \\
\multicolumn{2}{c} {} &
\multicolumn{1}{c} {$\lambda_{\rm RMBPT}$}&
\multicolumn{1}{c} {$\lambda_{\rm expt}$}&
\multicolumn{1}{c} {$A_r$} &
\multicolumn{1}{c} {$\lambda_{\rm RMBPT}$}&
\multicolumn{1}{c} {$\lambda_{\rm expt}$}&
\multicolumn{1}{c} {$A_r$} &
\multicolumn{1}{c} {$\lambda_{\rm RMBPT}$} &
\multicolumn{1}{c} {$\lambda_{\rm expt}$}&
\multicolumn{1}{c} {$A_r$} \\
\hline
$3s3s(^1S)2s\ ^2S_{1/2}$&$3p\ ^2P_{1/2}$& 11.407& 11.403&8.906[12]& 12.466& 12.461&7.918[12]&  10.475& 10.481&9.805[12]\\
$3d3d(^1S)2p\ ^2P_{1/2}$&$3d\ ^2D_{3/2}$& 11.425& 11.427&9.183[12]& 12.484& 12.492&7.808[12]&  10.494& 10.493&1.065[13]\\
$3d3d(^3P)2p\ ^2P_{3/2}$&$3d\ ^2D_{3/2}$& 11.488& 11.480&1.611[13]& 12.554& 12.556&1.383[13]&  10.551&       &1.819[13]\\
$3p3d(^3D)2p\ ^2S_{1/2}$&$3p\ ^2P_{3/2}$& 11.506& 11.503&3.173[13]& 12.572&       &2.698[13]&  10.569&       &3.654[13]\\
$3d3d(^3F)2p\ ^2D_{3/2}$&$3d\ ^2D_{5/2}$& 11.514&       &1.413[13]& 12.581&       &1.236[13]&  10.576& 10.573&1.566[13]\\
$3d3d(^1G)2p\ ^2F_{7/2}$&$3d\ ^2D_{5/2}$& 11.537& 11.538&2.019[13]& 12.592& 12.596&1.686[13]&  10.593&       &2.410[13]\\
$3p3d(^3P)2p\ ^2D_{3/2}$&$3p\ ^2P_{1/2}$& 11.552&       &2.305[13]& 12.624& 12.623&2.555[13]&  10.606&       &9.851[12]\\
$3p3d(^3P)2p\ ^4D_{3/2}$&$3p\ ^2P_{1/2}$& 11.560& 11.561&1.012[13]& 12.639&       &3.548[12]&  10.616&       &2.617[13]\\
$3d3d(^3F)2p\ ^2F_{5/2}$&$3d\ ^2D_{5/2}$& 11.615&       &3.112[12]& 12.694&       &2.364[12]&  10.664& 10.664&4.022[12]\\
$3p3d(^3P)2p\ ^2P_{3/2}$&$3p\ ^2P_{3/2}$& 11.624& 11.620&1.252[12]& 12.706& 12.700&9.215[11]&  10.674&       &1.493[12]\\
$3d3d(^3F)2p\ ^2D_{5/2}$&$3d\ ^2D_{5/2}$& 11.634&       &1.438[12]& 12.714&       &1.553[12]&  10.683& 10.687&1.858[12]\\
$3d3d(^3P)2p\ ^4S_{3/2}$&$3d\ ^2D_{5/2}$& 11.662&       &4.945[12]& 12.730& 12.736&3.862[12]&  10.723& 10.721&6.355[12]\\
$3p3d(^3D)2p\ ^4D_{5/2}$&$3p\ ^2P_{3/2}$& 11.673& 11.673&9.418[11]& 12.744&       &1.890[09]&  10.729&       &2.888[12]\\
$3s3d(^3D)2p\ ^4D_{3/2}$&$3s\ ^2S_{1/2}$& 11.688& 11.685&7.023[12]& 12.767&       &4.600[12]&  10.740& 10.738&1.021[13]\\
$3d3d(^3P)2p\ ^4D_{1/2}$&$3d\ ^2D_{3/2}$& 11.701&       &1.809[13]& 12.775& 12.772&1.361[13]&  10.758&       &2.344[13]\\
$3d3d(^3F)2p\ ^4F_{5/2}$&$3d\ ^2D_{3/2}$& 11.723&       &9.844[12]& 12.804&       &6.190[12]&  10.774& 10.778&1.268[13]\\
$3s3d(^3D)2p\ ^4D_{1/2}$&$3s\ ^2S_{1/2}$& 11.736& 11.737&7.629[12]& 12.822&       &6.064[12]&  10.782&       &9.682[12]\\
$3d3d(^1G)2p\ ^2G_{7/2}$&$3d\ ^2D_{5/2}$& 11.755& 11.755&7.229[12]& 12.836& 12.835&5.328[12]&  10.825&       &9.462[12]\\
$3p3d(^3D)2p\ ^4D_{1/2}$&$3p\ ^2P_{3/2}$& 11.785& 11.784&5.716[11]& 12.864&       &4.781[11]&  10.837&       &6.885[11]\\
\end{tabular}\end{ruledtabular}
\label{tab-com2}
\end{table*}

\begin{figure*}[tbp]
\centerline{\includegraphics[scale=0.35]{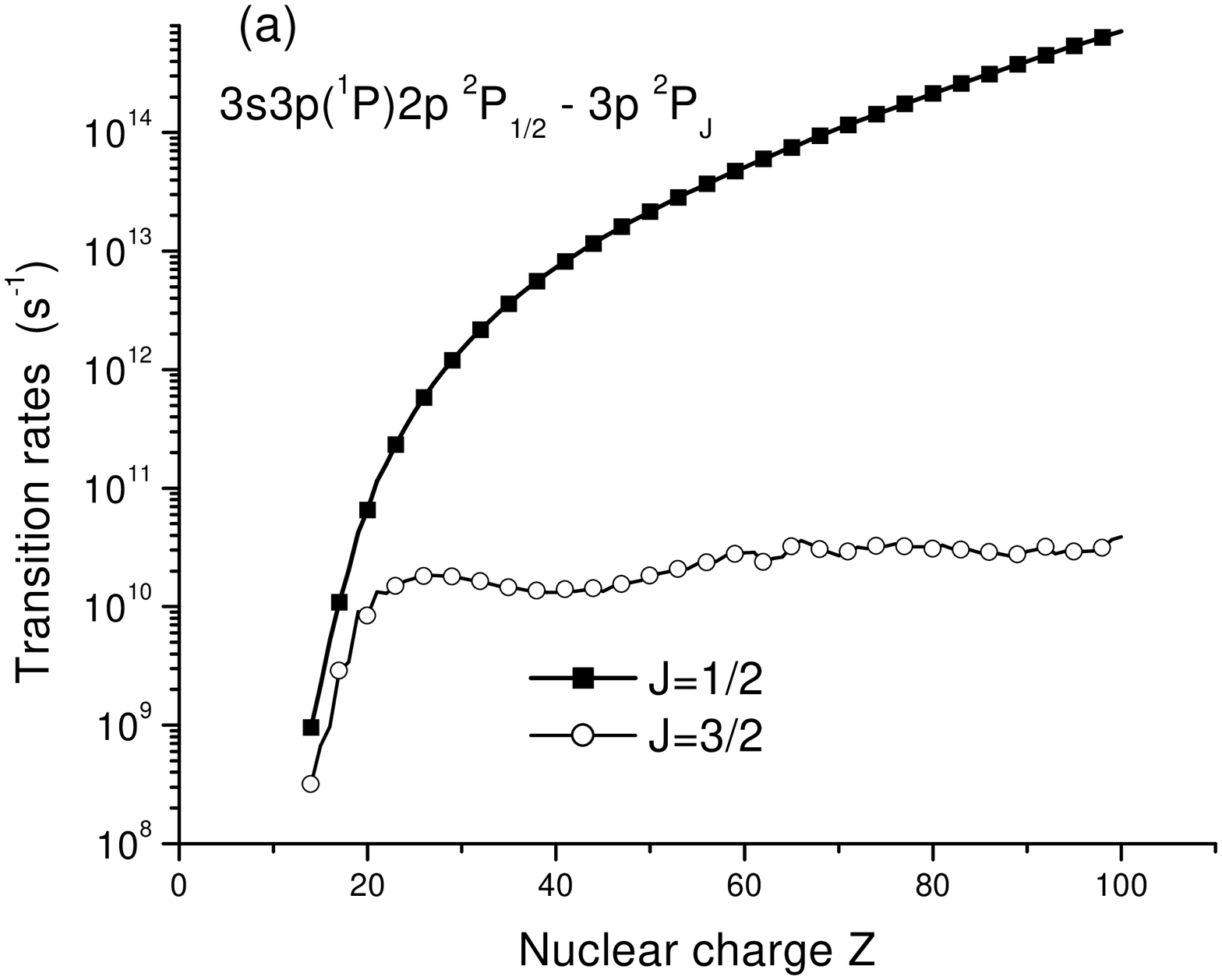}
            \includegraphics[scale=0.35]{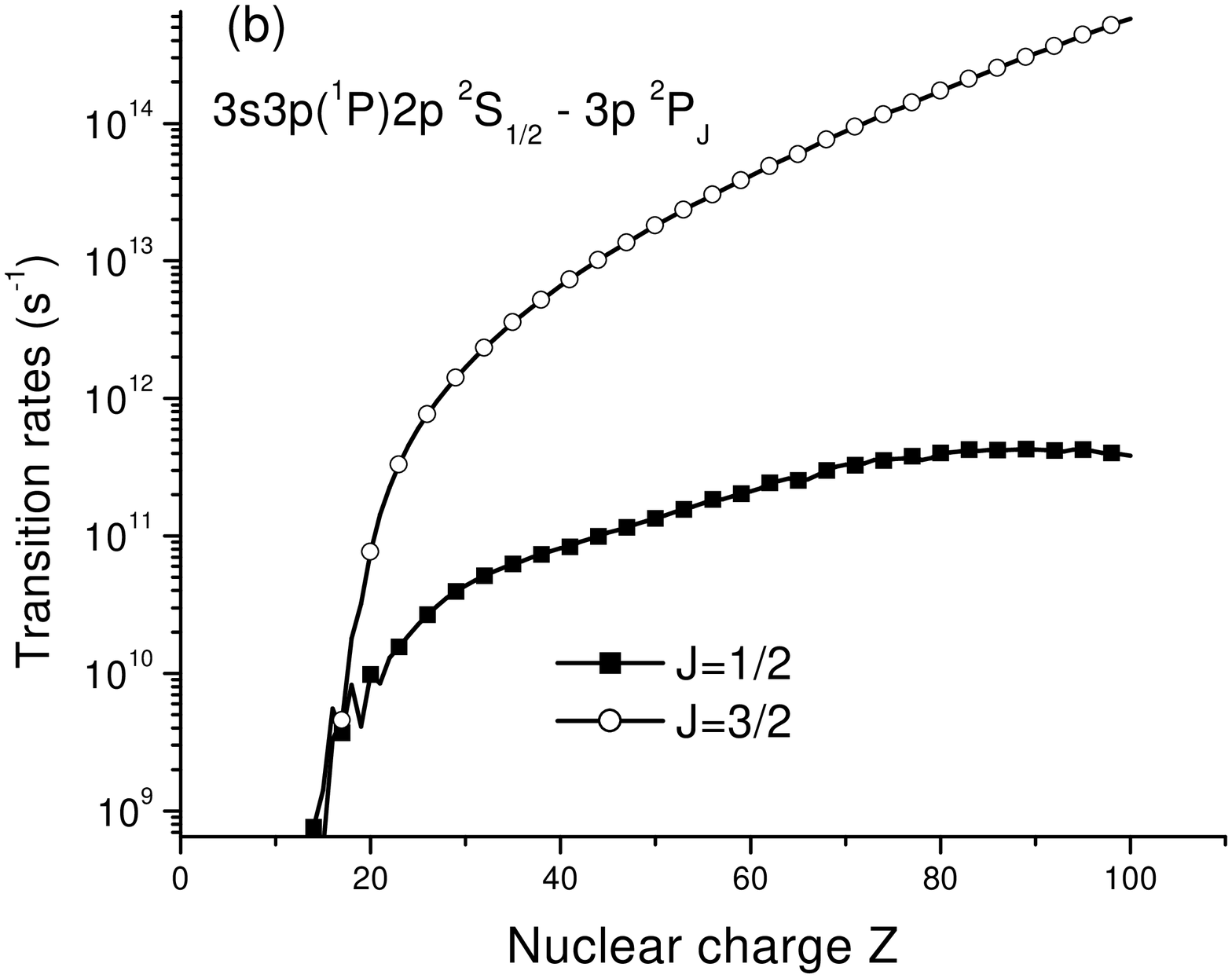}}
\centerline{\includegraphics[scale=0.35]{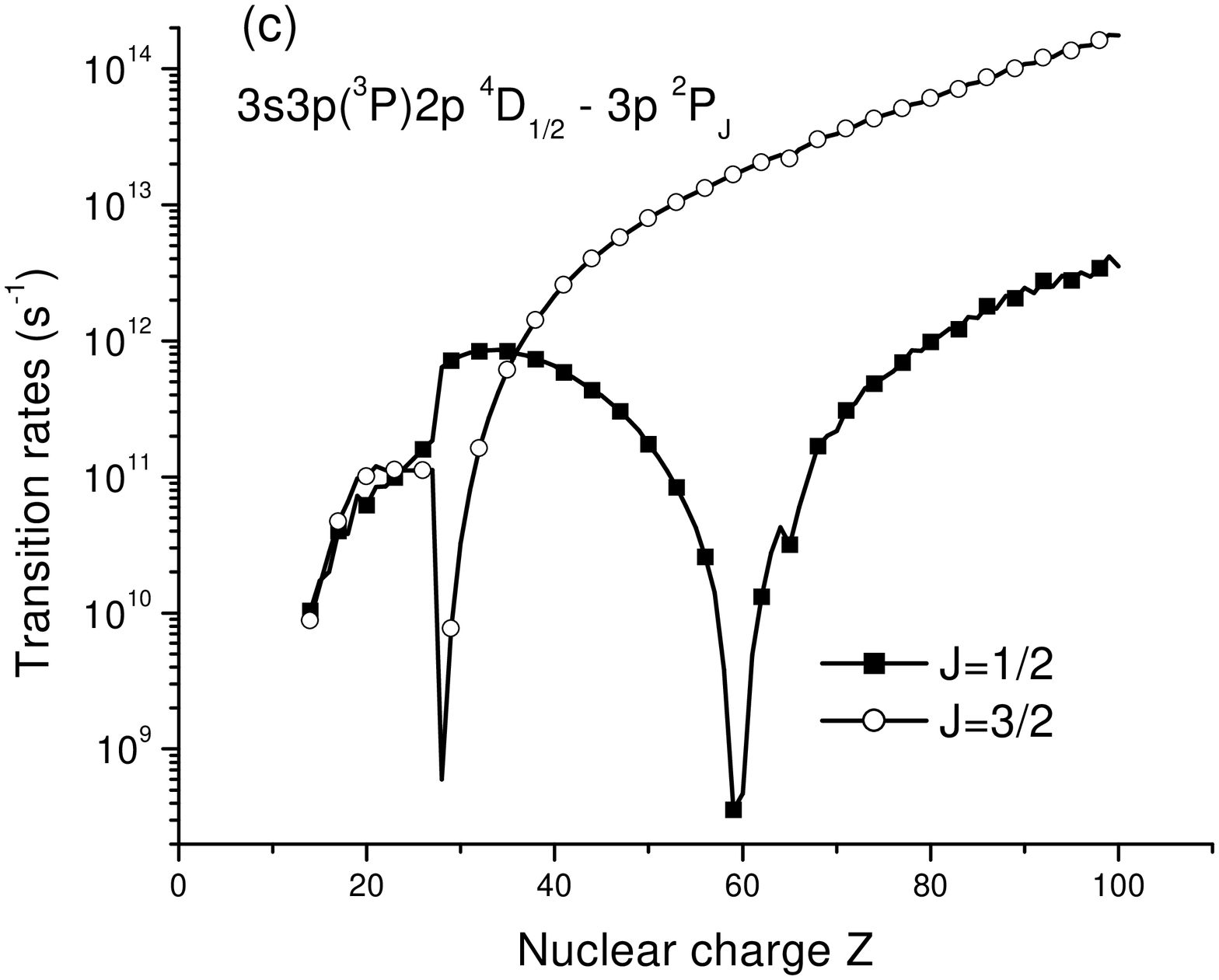}
            \includegraphics[scale=0.35]{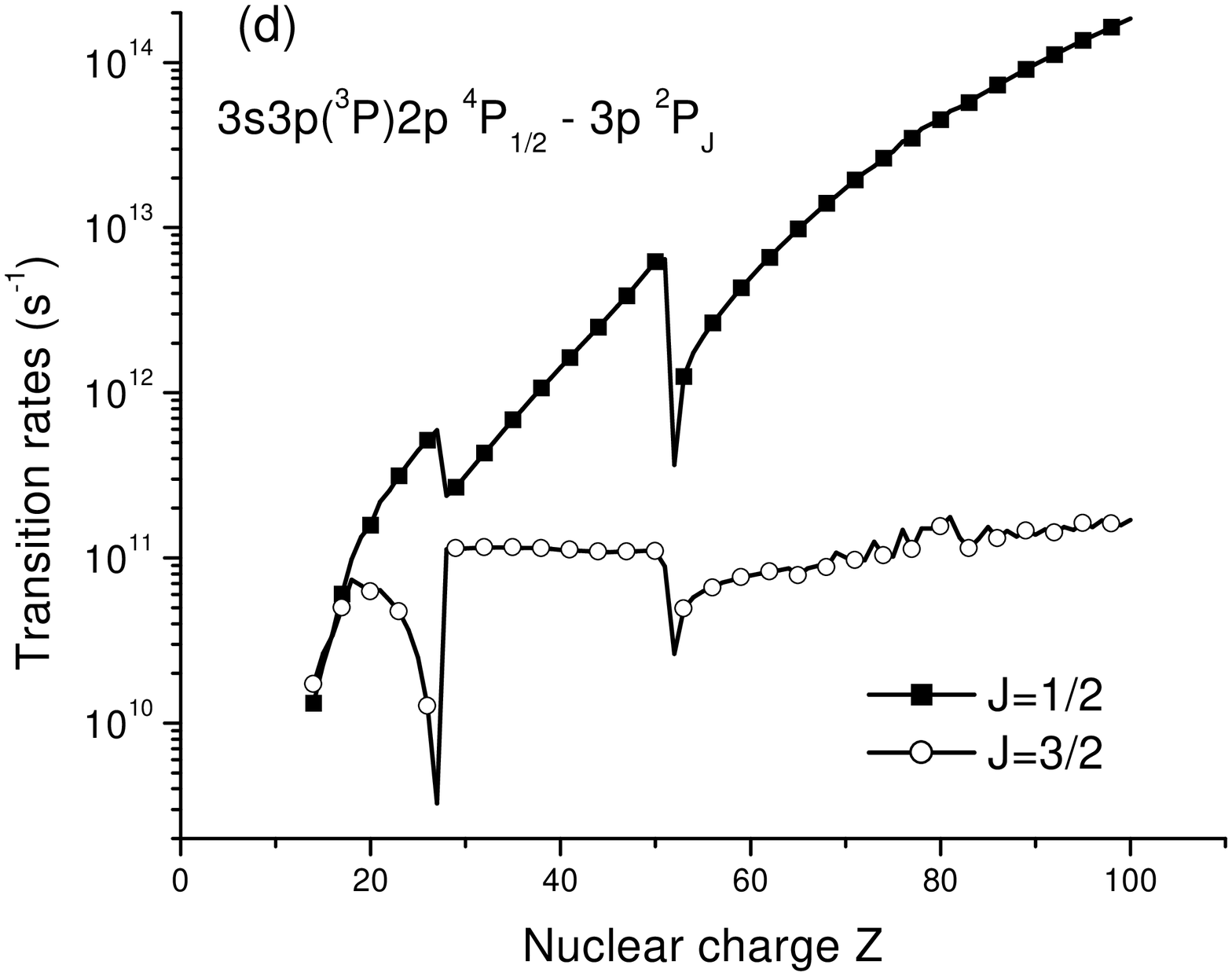}}
\centerline{\includegraphics[scale=0.35]{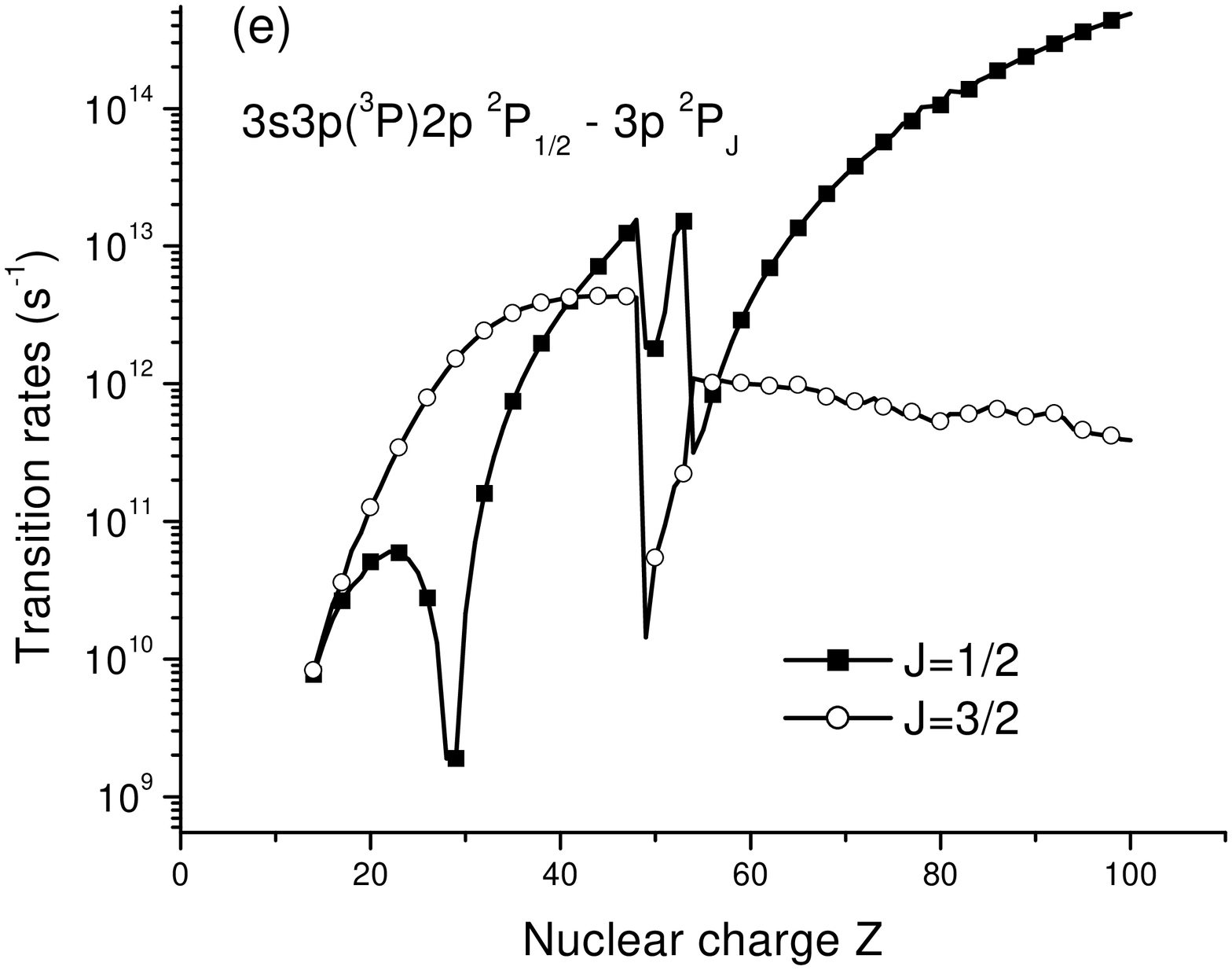}
            \includegraphics[scale=0.35]{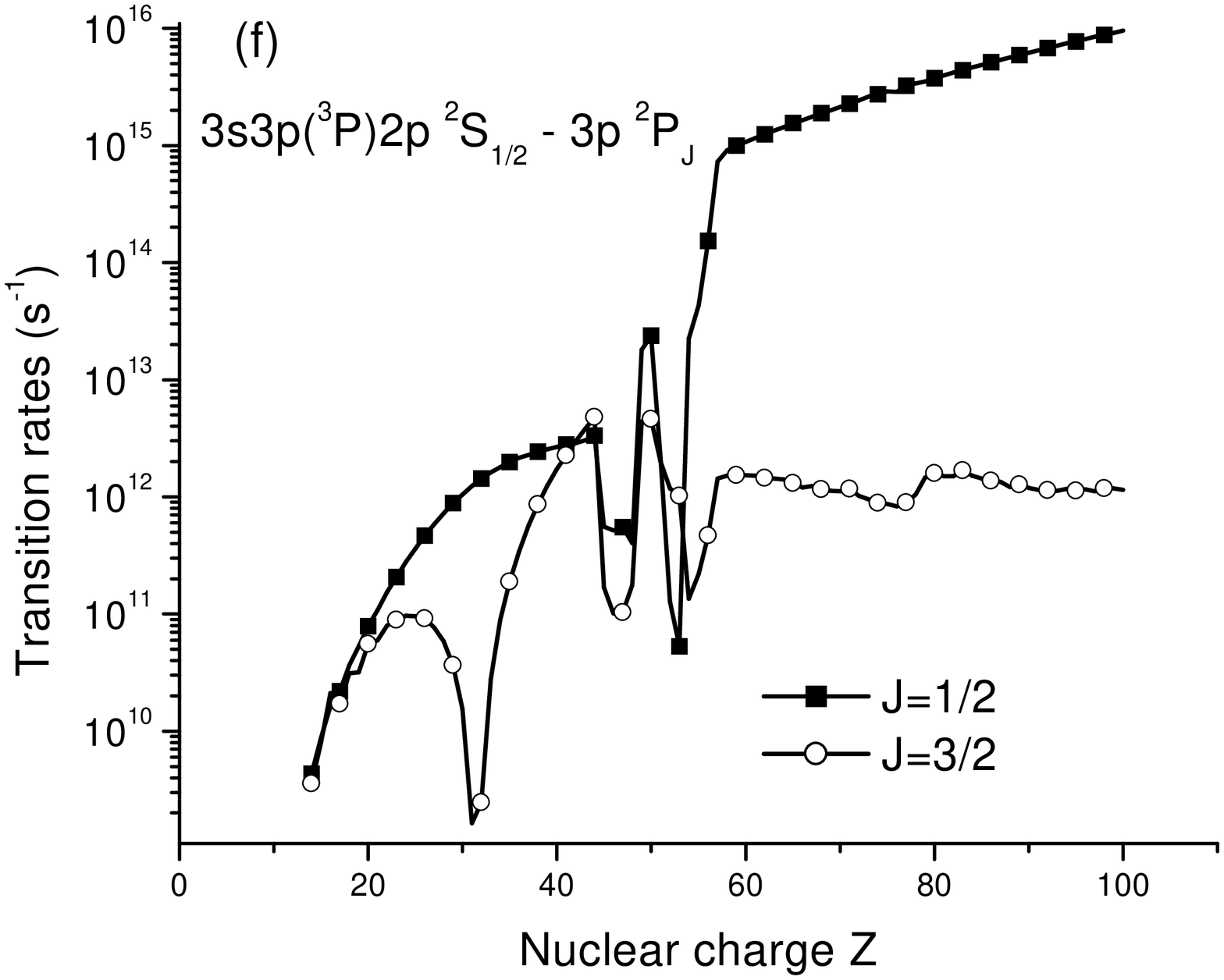}}
\caption{Transition rates for the transitions from core-excited even-parity states
with $J$ = 1/2 as function of $Z$ in Na-like ions.}
\label{fig-eve05}
\end{figure*}

\section{Results and comparison with other theory and experiment}

Trends of the $Z$-dependence of transition rates for
the transitions from core-excited even-parity
states with $J$ = 1/2 to two possible singly-excited odd-parity
states  are presented in Fig.~\ref{fig-eve05}.
Figures for transitions from other states are found in 
the accompanying EPAPS document \cite{EPAPS}.

 We find that transitions with smooth $Z$-dependence
are rarer than transitions with sharp features.  Smooth
 $Z$-dependence occurs for  transitions from doublet and quartet
 core-excited states. Usually, singularities occur in the
 intermediate interval of $Z$ = 25 - 50 when neither $LS$ nor $jj$ coupling
schemes describe the states of these ions properly.
  One general conclusion
 that can be derived from the figures is that the smooth
$Z$-dependence occurs more frequently for transitions from
low-lying core-excited states.

Singularities in the transition-rate curves have two distinct
origins: avoided level crossings and zeros in dipole matrix
elements. Avoided level crossings result in changes of the
dominant configuration of a state at a particular value of $Z$ and
lead to abrupt changes in the transition rate curves when the
partial rates associated with the dominant configurations below
and above the crossing point are significantly different. Zeros in
transition matrix elements lead to cusp-like
minima in the transition rate curves.  Examples of each of these
two singularity types can be seen in Fig.~\ref{fig-eve05}.

In Table~~\ref{tab-nils}, we present  data for transitions from odd-parity states
with $J=1/2$ in Fe$^{+15}$ and Xe$^{+43}$. We compare the present RMBPT
values with those given by \citet{m3}. More complete comparisons are given
in the accompanying EPAPS document \cite{EPAPS}. The calculations of \cite{m3} were based on
a multiconfigurational relativistic bound-state and distorted-wave continuum code
(YODA). Since $jj$ labeling was used in \cite{m3}, we keep that labeling in Table~\ref{tab-nils}.
 We find that the $A_r$-values from RMBPT and YODA differ by 10\% in most cases.
The differences are explained by 
the second-order corrections to dipole matrix elements
included in RMBPT.  

In Tables~\ref{tab-fe} - \ref{tab-com2}, wavelengths and
electric-dipole transition rates are presented for transitions in Na-like
Fe, Co, Ni, Cu and Zn.  We limit the tables to transitions
given in Refs.~\cite{a1,m94,brown}.
Measurements for Fe$^{15+}$ are presented in Tables ~\ref{tab-fe}
and  \ref{tab-com1} since two different ranges of spectra were
investigated in Ref.~\cite{a1} (16.8 - 17.8~\AA) and
Ref.~\cite{m94} (15.1 - 15.5 \AA). Three lines for Fe$^{15+}$
were identified  in region (15.1 - 15.26~\AA) by Brown {\it et
al.\/} in Ref.~\cite{brown}. All
possible $3l_1j_13l_2j_2[J_1]2l_3j_3 (J)$ - $3l_{j}$  transitions
produce 393 spectrum lines.  These lines in Fe$^{15+}$ are
covered by four spectral regions; 12.5 - 14.2~\AA~ (114 lines), 15.1
- 15.9~\AA~ (174 lines), 16.8 - 17.9~\AA~ (102 lines), and 19.3 -
19.7~\AA~ (3 lines). The first 114 lines are from
$3dj_13dj_2[J_1]2s (J)$ - $3l_{j}$  and
$3pj_13dj_2[J_1]2s_{1/2} (J)$ - $3l_{j}$ transitions and the last
three lines are from $3s3s[0]2p_{j} (J)$ - $3d_{j}$
transitions.
 Our RMBPT data together with experimental
measurements for Fe$^{15+}$ in the region of 16.8 - 17.8~\AA~ and
15.1 - 15.5~\AA~ are presented in Tables \ref{tab-fe} and
\ref{tab-com1}, respectively. 
  The agreement between our
RMBPT wavelengths and the experimental values is  0.02 - 0.04\% 
for both regions of the spectrum.

\section{Conclusion}

We have presented a systematic second-order relativistic MBPT
study of reduced matrix elements  and transition rates for
[$3l_1j_13l_2j_2[J_1]2l_3j_3\ (J)$ - $3lj$] electric-dipole
transitions in sodiumlike ions with the nuclear charges $Z$
ranging from 14 to 100. Our retarded $E1$ matrix elements include
correlation corrections from Coulomb and Breit interactions. Both
length and velocity forms of the matrix elements were evaluated
and small differences (0.4\% - 1\%), caused by the non locality of
the starting DF potential, were found between the two forms.
Second-order RMBPT transition energies were used in our evaluation
of transition rates. These calculations
were compared with other calculations and with available
experimental data. For $Z \geq 20$, we believe that the present
theoretical data are more accurate than other theoretical or
experimental data for transitions between the
$3l_1j_13l_2j_2[J_1]2l_3j_3\ (J)$ core-excited  states and the
$3lj$ singly-excited states in Na-like ions. We hope that these
results will be useful in analyzing older experiments and planning
new ones. Additionally, these calculations provide basic
theoretical input amplitudes for calculations of reduced matrix
elements, oscillator strengths and transition rates for Cu-like
satellites to transitions in Ni-like ions.

\begin{acknowledgments}
The work of W.R.J. and M.S.S. was supported in part by National
Science Foundation Grant No.\ PHY-0139928. U.I.S. acknowledges
partial support by Grant No.\ B516165 from Lawrence Livermore
National Laboratory. The work of J.R.A. was performed under the
auspices of the U. S. Department of Energy by the University of
California, Lawrence Livermore National Laboratory under contract
No.\ W-7405-Eng-48.
\end{acknowledgments}

\appendix*

\section{The particle-particle-hole diagram contribution for dipole matrix element}
\begin{widetext}

The first-order reduced E1 matrix element $Z^{(1)}$ for a
transition between the uncoupled particle-particle-hole
state $\Psi (QJM)$ of Eq.~(\ref{eq1}) and the one-particle state $%
a_{x}^{\dagger}|0\rangle $ is

\begin{equation}
\hspace{-0.5cm}Z^{(1)}(v^{0}w^{0}\left[ J_{12}\right]
aJ,xJ^{\prime
})=\sum_{vw}\sqrt{[J_{12}][J]}P_{J_{12}}(v^{0}v,w^{0}w)\delta
(J^{\prime }x)\delta (vx)Z(va)\left\{
\begin{array}{lll}
j_{x} & J & 1 \\
j_{a} & j_{w} & J_{12}
\end{array}
\right\} (-1)^{-j_{a}+j_{w}},
\end{equation}
where $v$, $w$ range over $\{v^0,w^0\}$.
 The quantity $P_{J}(v^{0}v,w^{0}w)$ is a symmetry
coefficient defined by
\begin{equation}
P_{J}(v^{0}v,w^{0}w)=\eta _{v^{0}w^{0}}\left[ \delta
_{v^{0}v}\delta _{w^{0}w}+(-1)^{j_{v}+j_{w}+J+1}\delta
_{v^{0}w}\delta _{w^{0}v}\right] , \label{eq4}
\end{equation}
where $\eta _{vw}$ is a normalization factor given by
\[
\eta _{vw}=\left\{
\begin{array}{ll}
1 & \mbox{for $w \neq v$} \\
1/\sqrt{2} & \mbox{for $w = v$}.
\end{array}
\right.
\]
The dipole matrix element $Z(va)$, which includes retardation, is given in
velocity and length forms in Eqs.(3,4) of Ref.~\cite{be-tr}. The
second-order reduced matrix element $Z^{(2)}(v^{0}w^{0}\left[ J_{12}\right]
aJ,xJ^{\prime })$ consists of four contributions: $Z^{({\rm HF})}$, $Z^{(%
{\rm RPA})}$, $Z^{({\rm corr})}$, and $Z^{({\rm derv})}$.
\begin{eqnarray}
&&\hspace{-0.4cm}Z^{(\mathrm{HF})}(v^{0}w^{0}\left[ J_{12}\right]
aJ,xJ^{\prime
})=\sum_{vw}\sqrt{[J_{12}][J]}P_{J_{12}}(v^{0}v,w^{0}w)\delta
(J^{\prime }x)\delta (vx)(-1)^{-j_{a}+j_{w}}  \nonumber  \label{eq-hf} \\
&&\hspace{-0.4cm}\times \left\{
\begin{array}{lll}
j_{x} & J & 1 \\
j_{a} & j_{w} & J_{12}
\end{array}
\right\} \sum_{i}\left[ \frac{\delta (j_{w}j_{i})\Delta
(wi)Z(ia)}{\epsilon (w)-\epsilon (i)}+\frac{\Delta (ia)\delta
(j_{a},j_{i})Z(wi)}{\epsilon (a)-\epsilon (i)}\right]   \nonumber
\end{eqnarray}

\begin{eqnarray}
Z^{({\rm RPA})}(v^{0}w^{0}\left[ J_{12}\right] aJ,xJ^{\prime
})&=&\frac{1}{3}\sum_{vw}\sqrt{[J_{12}][J]}P_{J_{12}}(v^{0}v,w^{0}w)
\delta (J^{\prime }x)\delta
(vx)\left\{
\begin{array}{lll}
j_{x} & J & 1 \\
j_{a} & j_{w} & J_{12}
\end{array}
\right\} (-1)^{-j_{a}+j_{w}} \nonumber \\
&\times&\sum_{nb}\left[
\frac{Z_{1}(wban)Z(bn)}{\epsilon (b)+\epsilon (w)-\epsilon
(a)-\epsilon (n)}+\frac{Z_{1}(wnab)Z(nb)}{\epsilon (b)+\epsilon
(a)-\epsilon (w)-\epsilon (n)}\right] \label{eq-rpa}
\end{eqnarray}

\begin{eqnarray}
Z^{({\rm corr})}(v^{0}w^{0}\left[ J_{12}\right] aJ,xJ^{\prime })
&=&\sum_{vw}\sqrt{[J_{12}][J]}P_{J_{12}}(v^{0}v,w^{0}w)\delta
(J^{\prime
}x)\sum_{k}(-1)^{j_{a}-j_{v}+J_{12}+k}  \nonumber \\
&\times&\sum_{i}\left[ \frac{Z(ix)X_{k}(vwai)}{\epsilon (v)+\epsilon (w)-\epsilon
(a)-\epsilon (i)}\delta (J,j_{i})\frac{1}{[J]}\left\{
\begin{array}{lll}
j_{w} & j_{i} & k \\
j_{a} & j_{v} & J_{12}
\end{array}
\right\} \right.   \nonumber \\
&-&\left. \frac{Z(ia)X_{k}(vwxi)}{\epsilon (v)+\epsilon (w)-\epsilon
(x)-\epsilon (i)}\left\{
\begin{array}{lll}
j_{x} & j_{i} & J_{12} \\
j_{w} & j_{v} & k
\end{array}
\right\} \left\{
\begin{array}{lll}
j_{x} & J & 1 \\
j_{a} & j_{i} & J_{12}
\end{array}
\right\} (-1)^{J+j_{w}-j_{i}}\right.   \nonumber \\
&-&\left. \frac{Z(vi)Z_{k}(axwi)}{\epsilon (a)+\epsilon (x)-\epsilon
(w)-\epsilon (i)}\left\{
\begin{array}{lll}
j_{v} & J & k \\
j_{a} & j_{w} & J_{12}
\end{array}
\right\} \left\{
\begin{array}{lll}
j_{x} & J & 1 \\
j_{v} & j_{i} & k
\end{array}
\right\} (-1)^{k+J_{12}+j_{w}+j_{i}-J+1}\right]  \label{eq-corr}
\end{eqnarray}

 In the above equations, the index $b$ designates
core states, index $n$ designates excited states, and index $i$ denotes an
arbitrary core or excited state. In the sums over $i$ in
Eqs.~(\ref{eq-hf},\ref{eq-corr}), all terms with vanishing
denominators are excluded. The definitions of $%
X_{k}(abcd)$ and $Z_{k}(abcd)$ are given by Eq.(2.12) and Eq.(2.15) in Ref.~%
\cite{be-en} and $\Delta _{ij}$ is defined at the end of section II in \cite
{be-en}; $\epsilon (w)$ is a one-electron DF energy.

The derivative term is just the derivative of the the first-order matrix
element with respect to the transition energy. It is introduced to account
for the first-order change in transition energy. An auxiliary quantity $P^{(%
{\rm derv})}$ is defined by

\begin{equation}
P^{({\rm derv})}(v^{0}w^{0}\left[ J_{12}\right] aJ,xJ^{\prime })=\sum_{vw}%
\sqrt{[J_{12}][J]}P_{J_{12}}(v^{0}v,w^{0}w)
\delta (J^{\prime }x)\delta (vx)Z^{({\rm derv})}(va)\left\{
\begin{array}{lll}
j_{x} & J & 1 \\
j_{a} & j_{w} & J_{12}
\end{array}
\right\} (-1)^{-j_{a}+j_{w}}.
\end{equation}

The derivative term $Z^{({\rm derv)}}(va)$ is given in length and velocity
forms by Eqs.~(10) and (11) of Ref.~\cite{be-tr}.

The coupled dipole transition matrix element between the initial state $I$
 and final state $F$ in Na-like ions is given by

\begin{eqnarray}
Q^{(1+2)}(I-F)&=&-\frac{1}{E^{(1)}[F]-E^{(1)}[I]}\sum_{avw}%
\sum_{J_{12}^{\prime }}
C_{1}^{F}[vw[J_{12}^{\prime }]a(J)]  \nonumber \\
&\times& \left\{ \left[ \epsilon(x)-\epsilon(vwa)\right] \left[ Z^{(1+2%
{\rm )}}[vw\left[ J_{12}^{\prime }\right] aJ,xJ^{\prime
}]+B^{(2)}[vw\left[
J_{12}^{\prime }\right] aJ,xJ^{\prime }]\right] \right.  \nonumber \\
 &+& \left.\left[ -E^{(1)}[F]+E^{(1)}[I]-\epsilon(x)+\epsilon(vwa)\right]
 P^{({\rm derv)}}[vw\left[ J_{12}^{\prime }\right]
aJ,xJ^{\prime }]\right\}. \label{coupl}
\end{eqnarray}

Here, $\epsilon(vwa)=\epsilon(v)+\epsilon(w)-\epsilon(a)$, $%
Z^{(1+2)}=Z^{(1)}+Z^{({\rm RPA})}+Z^{({\rm corr})}$. (Note that $Z^{({\rm HF}%
)}$ vanishes since we start from a Hartree-Fock basis.)
The sum over $vwa,\,J_{12}^{\prime}$ is understood as sum over
the complex of states with the same $J$ and parity.
 In Eq.~(\ref{coupl}), we let $%
B^{(2)}=B^{({\rm RPA})}+B^{({\rm HF)}}$ +$B^{({\rm corr})}$ to represent
second-order corrections arising from the Breit interaction. The quantities $%
C_{1}^{F}[vw[J_{12}^{\prime }]a(J)]$ are eigenvectors (or mixing
coefficients) for particle-particle-hole state $F$. The initial
state $I$ is a single, one-valence state. Using the above
formulas and the results for uncoupled reduced matrix elements, we
carry out the transformation from uncoupled reduced matrix elements to intermediate
coupled matrix elements between physical states.
\end{widetext}


\end{document}